\documentclass[a4paper,11pt]{article}
\pdfoutput=1 
\usepackage[T1]{fontenc} 

\makeatletter
\gdef\@fpheader{ }
\gdef\@journal{ }
\makeatother
\RequirePackage{amsmath}
\RequirePackage{amssymb}
\RequirePackage{epsfig}
\RequirePackage{graphicx}
\RequirePackage[numbers,sort&compress]{natbib}
\RequirePackage{color}
\RequirePackage[colorlinks=true
,urlcolor=blue
,anchorcolor=blue
,citecolor=blue
,filecolor=blue
,linkcolor=blue
,menucolor=blue
,pagecolor=blue
,linktocpage=true
,pdfproducer=medialab
,pdfa=true
]{hyperref}

\newif\ifnotoc\notocfalse
\newif\ifemailadd\emailaddfalse
\newif\iftoccontinuous\toccontinuousfalse
\makeatletter
\def\@subheader{\@empty}
\def\@keywords{\@empty}
\def\@abstract{\@empty}
\def\@xtum{\@empty}
\def\@dedicated{\@empty}
\def\@arxivnumber{\@empty}
\def\@collaboration{\@empty}
\def\@collaborationImg{\@empty}
\def\@proceeding{\@empty}
\def\@preprint{\@empty}

\newcommand{\subheader}[1]{\gdef\@subheader{#1}}
\newcommand{\keywords}[1]{\if!\@keywords!\gdef\@keywords{#1}\else%
\PackageWarningNoLine{\jname}{Keywords already defined.\MessageBreak Ignoring last definition.}\fi}
\renewcommand{\abstract}[1]{\gdef\@abstract{#1}}
\newcommand{\dedicated}[1]{\gdef\@dedicated{#1}}
\newcommand{\arxivnumber}[1]{\gdef\@arxivnumber{#1}}
\newcommand{\proceeding}[1]{\gdef\@proceeding{#1}}
\newcommand{\xtumfont}[1]{\textsc{#1}}
\newcommand{\correctionref}[3]{\gdef\@xtum{\xtumfont{#1} \href{#2}{#3}}}
\newcommand\jname{JHEP}
\newcommand\acknowledgments{\section*{Acknowledgments}}

\newcommand\preprint[1]{\gdef\@preprint{\hfill #1}}

\makeatother

\newcommand\note[2][]{%
\if!#1!%
\stepcounter{footnote}\footnotetext{#2}%
\else%
{\renewcommand\thefootnote{#1}%
\footnotetext{#2}}%
\fi}

\makeatletter
\newtoks\auth@toks
\renewcommand{\author}[2][]{%
  \if!#1!%
    \auth@toks=\expandafter{\the\auth@toks#2\ }%
  \else
    \auth@toks=\expandafter{\the\auth@toks#2$^{#1}$\ }%
  \fi
}
\makeatother
\makeatletter
\newtoks\affil@toks\newif\ifaffil\affilfalse
\newcommand{\affiliation}[2][]{%
\affiltrue
  \if!#1!%
    \affil@toks=\expandafter{\the\affil@toks{\item[]#2}}%
  \else
    \affil@toks=\expandafter{\the\affil@toks{\item[$^{#1}$]#2}}%
  \fi
}
\makeatother
\makeatletter
\newtoks\email@toks\newcounter{email@counter}%
\newcommand{\emailAdd}[1]{%
\emailaddtrue%
\ifnum\theemail@counter>0\email@toks=\expandafter{\the\email@toks, \@email{#1}}%
\else\email@toks=\expandafter{\the\email@toks\@email{#1}}%
\fi\stepcounter{email@counter}}
\newcommand{\@email}[1]{\href{mailto:#1}{\tt #1}}
\makeatother

\makeatletter
\newcommand*\collaboration[1]{\gdef\@collaboration{#1}}
\newcommand*\collaborationImg[2][]{\gdef\@collaborationImg{#2}}
\makeatletter
\newcommand\afterLogoSpace{\smallskip}
\newcommand\afterSubheaderSpace{\vskip3pt plus 2pt minus 1pt}
\newcommand\afterProceedingsSpace{\vskip21pt plus0.4fil minus15pt}
\newcommand\afterTitleSpace{\vskip23pt plus0.06fil minus13pt}
\newcommand\afterRuleSpace{\vskip23pt plus0.06fil minus13pt}
\newcommand\afterCollaborationSpace{\vskip3pt plus 2pt minus 1pt}
\newcommand\afterCollaborationImgSpace{\vskip3pt plus 2pt minus 1pt}
\newcommand\afterAuthorSpace{\vskip5pt plus4pt minus4pt}
\newcommand\afterAffiliationSpace{\vskip3pt plus3pt}
\newcommand\afterEmailSpace{\vskip16pt plus9pt minus10pt\filbreak}
\newcommand\afterXtumSpace{\par\bigskip}
\newcommand\afterAbstractSpace{\vskip16pt plus9pt minus13pt}
\newcommand\afterKeywordsSpace{\vskip16pt plus9pt minus13pt}
\newcommand\afterArxivSpace{\vskip3pt plus0.01fil minus10pt}
\newcommand\afterDedicatedSpace{\vskip0pt plus0.01fil}
\newcommand\afterTocSpace{\bigskip\medskip}
\newcommand\afterTocRuleSpace{\bigskip\bigskip}
\newlength{\affiliationsSep}
\newcommand\beforetochook{\pagestyle{myplain}\pagenumbering{roman}}

\DeclareFixedFont\trfont{OT1}{phv}{b}{sc}{11}

\renewcommand\maketitle{
\pagestyle{empty}
\thispagestyle{titlepage}
\setcounter{page}{0}
\noindent{\small\scshape\@fpheader}\@preprint\par

\afterLogoSpace
\if!\@subheader!\else\noindent{\trfont{\@subheader}}\fi
\afterSubheaderSpace
\if!\@proceeding!\else\noindent{\sc\@proceeding}\fi
\afterProceedingsSpace
{\LARGE\flushleft\sffamily\bfseries\@title\par}
\afterTitleSpace
\hrule height 1.5\p@%
\afterRuleSpace
\if!\@collaboration!\else
{\Large\bfseries\sffamily\raggedright\@collaboration}\par
\afterCollaborationSpace
\fi
\if!\@collaborationImg!\else
{\normalsize\bfseries\sffamily\raggedright\@collaborationImg}\par
\afterCollaborationImgSpace
\fi
{\bfseries\raggedright\sffamily\the\auth@toks\par}
\afterAuthorSpace
\ifaffil\begin{list}{}{%
\setlength{\leftmargin}{0.28cm}%
\setlength{\labelsep}{0pt}%
\setlength{\itemsep}{\affiliationsSep}%
\setlength{\topsep}{-\parskip}}
\itshape\small%
\the\affil@toks
\end{list}\fi
\afterAffiliationSpace
\ifemailadd 
\noindent\hspace{0.28cm}\begin{minipage}[l]{.9\textwidth}
\begin{flushleft}
\textit{E-mail:} \the\email@toks
\end{flushleft}
\end{minipage}
\else 
\PackageWarningNoLine{\jname}{E-mails are missing.\MessageBreak Plese use \protect\emailAdd\space macro to provide e-mails.}
\fi
\afterEmailSpace
\if!\@xtum!\else\noindent{\@xtum}\afterXtumSpace\fi
\if!\@abstract!\else\noindent{\renewcommand\baselinestretch{.9}\textsc{Abstract:}}\ \@abstract\afterAbstractSpace\fi
\if!\@keywords!\else\noindent{\textsc{Keywords:}} \@keywords\afterKeywordsSpace\fi
\if!\@arxivnumber!\else\noindent{\textsc{ArXiv ePrint:}} \href{http://arxiv.org/abs/\@arxivnumber}{\@arxivnumber}\afterArxivSpace\fi
\if!\@dedicated!\else\vbox{\small\it\raggedleft\@dedicated}\afterDedicatedSpace\fi
\ifnotoc\else
\iftoccontinuous\else\newpage\fi
\beforetochook\hrule
\tableofcontents
\afterTocSpace
\hrule
\afterTocRuleSpace
\fi
\setcounter{footnote}{0}
\pagestyle{myplain}\pagenumbering{arabic}
} 

\renewcommand{\baselinestretch}{1.1}\normalsize
\divide\@tempdima\baselineskip
\@tempcnta=\@tempdima
\skip\@mpfootins = \skip\footins
\renewcommand{\@dotsep}{10000}

\newcommand\ps@myplain{
\pagenumbering{arabic}
\renewcommand\@oddfoot{\hfill-- \thepage\ --\hfill}
\renewcommand\@oddhead{}}
\let\ps@plain=\ps@myplain

\newcommand\ps@titlepage{\renewcommand\@oddfoot{}\renewcommand\@oddhead{}}

\numberwithin{equation}{section}

\renewcommand\section{\@startsection{section}{1}{\z@}%
                                   {-3.5ex \@plus -1.3ex \@minus -.7ex}%
                                   {2.3ex \@plus.4ex \@minus .4ex}%
                                   {\normalfont\large\bfseries}}
\renewcommand\subsection{\@startsection{subsection}{2}{\z@}%
                                   {-2.3ex\@plus -1ex \@minus -.5ex}%
                                   {1.2ex \@plus .3ex \@minus .3ex}%
                                   {\normalfont\normalsize\bfseries}}
\renewcommand\subsubsection{\@startsection{subsubsection}{3}{\z@}%
                                   {-2.3ex\@plus -1ex \@minus -.5ex}%
                                   {1ex \@plus .2ex \@minus .2ex}%
                                   {\normalfont\normalsize\bfseries}}
\renewcommand\paragraph{\@startsection{paragraph}{4}{\z@}%
                                   {1.75ex \@plus1ex \@minus.2ex}%
                                   {-1em}%
                                   {\normalfont\normalsize\bfseries}}
\renewcommand\subparagraph{\@startsection{subparagraph}{5}{\parindent}%
                                   {1.75ex \@plus1ex \@minus .2ex}%
                                   {-1em}%
                                   {\normalfont\normalsize\bfseries}}

\def\fnum@figure{\textbf{\figurename\nobreakspace\thefigure}}
\def\fnum@table{\textbf{\tablename\nobreakspace\thetable}}

\long\def\@makecaption#1#2{%
  \vskip\abovecaptionskip
  \sbox\@tempboxa{\small #1. #2}%
  \ifdim \wd\@tempboxa >\hsize
    \small #1. #2\par
  \else
    \global \@minipagefalse
    \hb@xt@\hsize{\hfil\box\@tempboxa\hfil}%
  \fi
  \vskip\belowcaptionskip}

\renewenvironment{thebibliography}[1]{%
\begin{oldthebibliography}{#1}%
\small%
\raggedright%
\setlength{\itemsep}{5pt plus 0.2ex minus 0.05ex}%
}%
{%
\end{oldthebibliography}%
}

\begin{document}


\title{\boldmath Covariant perturbation expansion of off-diagonal heat kernel}

\author[a]{Yu-Zi Gou,}
\author[a]{Wen-Du Li,}
\author[b,1]{Ping Zhang,}\note{zhangping@cueb.edu.cn.}
\author[a,c,2]{and Wu-Sheng Dai}\note{daiwusheng@tju.edu.cn.}


\affiliation[a]{Department of Physics, Tianjin University, Tianjin 300072, P.R. China}
\affiliation[b]{School of Finance, Capital University of Economics and Business, Beijing 100070, P. R. China}
\affiliation[c]{LiuHui Center for Applied Mathematics, Nankai University \& Tianjin University, Tianjin 300072, P.R. China}









\abstract{
Covariant perturbation expansion is an important method in quantum field
theory. In this paper an expansion up to arbitrary order for\ off-diagonal
heat kernels in flat space based on the covariant perturbation expansion is
given. In literature, only diagonal heat kernels are calculated based on the
covariant perturbation expansion.}

\maketitle
\flushbottom


\section{Introduction}

The heat-kernel method is widely used in physics
\cite{vassilevich2003heat,kirsten2001spectral} and mathematics
\cite{vassilevich2003heat,davies1990heat}. The heat-kernel method can extract
the information embedded in an operator; in physics the operator is often the
Hamiltonian \cite{vassilevich2003heat,kirsten2001spectral}. In quantum field
theory, for example, one can calculate one-loop effective actions, vacuum
energies, and other spectral functions by the heat-kernel method
\cite{dai2010approach,mukhanov2007introduction,vassilevich2003heat}. In
statistical mechanics, the heat-kernel method is very useful since the global
heat kernel, the trace of the heat kernel, is indeed the partition function
\cite{dai2003quantum,dai2004geometry,dai2009number}. In scattering theory, the
heat-kernel method can be used to calculate scattering phase shifts
\cite{li2015heat,pang2012relation}.

Developing various asymptotic expansions is a core issue in the study of the
heat-kernel method. There is much research on the perturbation calculation of
heat kernels. Nevertheless, most literature dealing with only diagonal heat
kernel
\cite{vassilevich2003heat,fliegner1994higher,fliegner1998higher,nepomechie1985calculating}%
, while there are only a very few literature calculates off-diagonal heat
kernel
\cite{culumovic1988calculation,dilkes1996off,mckeon1991seeley,kotani2000albanese,groh2011off}%
. Nevertheless, rather than the diagonal one, the off-diagonal heat kernel
contains all information of an operator, and many problems need off-diagonal,
e.g., the heat-kernel method for scattering \cite{li2015heat,pang2012relation}.

The covariant perturbation expansion, among various heat-kernel algorithms, is
a very effective method
\cite{barvinsky1985generalized,barvinsky1987beyond,barvinsky1995one,shore2002local,barvinsky1993covariant,barvinsky1994asymptotic,barvinsky1990covariant2,barvinsky1990covariant3,gusev2009heat,barvinsky2002new,barvinsky2010schwinger}%
. It can account for non-local effects \cite{van1998index}. Particularly, the
covariant perturbation expansion is uniformly convergent
\cite{barvinsky1987beyond,gusev2009heat}. The special advantage of the
covariant perturbation, as an example, is shown in Ref. \cite{li2015heat}. It
is provided in Ref. \cite{li2015heat} a heat-kernel method to calculate
scattering phase shifts. By this method, each method of calculating heat
kernels can be converted into a method of solving a scattering problem. It is
shown that the perturbation expansion of phase shift obtained by the covariant
perturbation expansion is more accurate than the Born approximation.

In a series of papers, only the diagonal heat kernel is considered
\cite{mukhanov2007introduction}. In this paper, we expand the off-diagonal
heat kernel to arbitrary order for flat space cases by the covariant
perturbation theory.

In section \ref{CPE}, we provide an expansion of heat kernels based on the
covariant perturbation expansion. In section \ref{Matrix}, we give an
expansion of off-diagonal heat kernels. In section \ref{alternative}, we give
an alternative expression of off-diagonal heat kernels. In section \ref{3},
the first three orders of diagonal and global heat kernels are given. The
conclusion is given in section \ref{Conclusion}.

\section{The covariant perturbation expansion \label{CPE}}

\subsection{The heat kernel}

For an operator $D$, we can define the heat kernel operator
\cite{vassilevich2003heat},%
\begin{equation}
K\left(  \tau\right)  =e^{-D\tau}. \label{1.2.1}%
\end{equation}
The heat kernel operator is determined by the heat equation%
\begin{equation}
\frac{d}{d\tau}K\left(  \tau\right)  +DK\left(  \tau\right)  =0\text{\ }
\label{1.2.2}%
\end{equation}
with the initial condition $K\left(  0\right)  =1$.

The heat kernel is the matrix element of the heat kernel operator
\cite{mukhanov2007introduction}
\begin{equation}
K\left(  \tau;x,y\right)  =\left\langle x\left\vert K\left(  \tau\right)
\right\vert y\right\rangle =\sum_{\lambda}e^{-\lambda\tau}\phi_{\lambda
}\left(  x\right)  \phi_{\lambda}^{\ast}\left(  y\right)  \label{1.2.3}%
\end{equation}
and the global heat kernel is the trace of the heat kernel
\begin{equation}
K_{D}\left(  \tau\right)  =\operatorname{tr}K\left(  \tau\right)  =\int
dxK\left(  \tau;x,x\right)  =\sum_{\lambda}e^{-\lambda\tau}, \label{1.1.5}%
\end{equation}
where $\phi_{\lambda}\left(  x\right)  $ and $\lambda$ are the eigenfunction
and eigenvalue of the operator $D$, determined by the eigenequation
$D\phi_{\lambda}=\lambda\phi_{\lambda}$.

\subsection{The covariant perturbation expansion}

In Refs.
\cite{mukhanov2007introduction,barvinsky1987beyond,barvinsky1990covariant2,barvinsky1990covariant3}%
, the covariant perturbation expansion is applied to calculate diagonal heat
kernels to first three orders. In Ref. \cite{li2015heat}, for calculating the
scattering phase shift, we use the covariant perturbation expansion to
calculate the non-diagonal heat kernel. In this paper, we give an expansion
of\ non-diagonal heat kernels to $n$-th order by the covariant perturbation method.

The expansion of a heat kernel operator in flat space can be expressed as
\cite{mukhanov2007introduction}%

\begin{equation}
K_{D}\left(  \tau\right)  =K_{0}\left(  \tau\right)  +K_{1}\left(
\tau\right)  +K_{2}\left(  \tau\right)  +\cdots, \label{1.4.1}%
\end{equation}
where $K_{n}\left(  \tau\right)  $ is the $n$-th order heat kernel operator.
In our cases,
\begin{equation}
D=-\nabla^{2}+V. \label{1.4.2}%
\end{equation}

In order to achieve the non-diagonal heat kernel, from Eqs. (\ref{1.2.2}) and
(\ref{1.4.1}), we achieve the following recurrence relations:%

\begin{equation}
\left\{
\begin{array}
[c]{c}%
\frac{dK_{0}}{d\tau}=\nabla^{2}K_{0},\text{ }K_{0}\left(  0\right)  =1\\
\frac{dK_{n}}{d\tau}=\nabla^{2}K_{n}-VK_{n-1},\text{ \ }K_{n}\left(  0\right)
=0
\end{array}
\right.  . \label{2.2}%
\end{equation}
The solution of Eq. (\ref{2.2}) can be expressed formally as%
\begin{align}
K_{0}\left(  \tau\right)   &  =\exp\left(  \tau\nabla^{2}\right)
,\label{2.3}\\
K_{n}\left(  \tau\right)   &  =K_{0}\left(  \tau\right)  C_{n}\left(
\tau\right)  . \label{2.3b}%
\end{align}
Note that $C_{n}\left(  \tau\right)  $ here is an operator.

Derivating Eq. (\ref{2.3}) with respect to $\tau$ gives%
\begin{equation}
\frac{d}{d\tau}K_{n}\left(  \tau\right)  =\nabla^{2}K_{0}C_{n}\left(
\tau\right)  +K_{0}\left(  \tau\right)  \frac{d}{d\tau}C_{n}\left(
\tau\right)  . \label{2.4}%
\end{equation}
From Eqs. (\ref{2.2}) and (\ref{2.4}), we obtain%
\begin{equation}
K_{0}\left(  \tau\right)  \frac{d}{d\tau}C_{n}\left(  \tau\right)
=-VK_{n-1}\left(  \tau\right)  . \label{2.5}%
\end{equation}
By integrating both sides of Eq. (\ref{2.5}) with respect to $\tau$, we obtain%
\begin{equation}
C_{n}\left(  \tau\right)  =\int_{0}^{\tau}d\tau_{n}K_{0}^{-1}\left(  \tau
_{n}\right)  \left(  -V\right)  K_{n-1}(\tau_{n}), \label{2.6}%
\end{equation}
where $C_{n}\left(  0\right)  =0$.

By Eq. (\ref{2.3}), we obtain
\begin{align}
K_{0}\left(  \tau\right)  K_{0}\left(  \tau_{1}\right)   &  =K_{0}\left(
\tau+\tau_{1}\right)  ,\text{ }\tau>0,\text{ }\tau_{1}>0,\label{2.7}\\
K_{0}^{-1}\left(  \tau\right)   &  =K_{0}\left(  -\tau\right)  . \label{2.8}%
\end{align}
By Eqs. (\ref{2.6}), (\ref{2.7}), and (\ref{2.8}), we arrive at%
\begin{equation}
K_{n}\left(  \tau\right)  =\int_{0}^{\tau}d\tau_{n}K_{0}\left(  \tau-\tau
_{n}\right)  \left(  -V\right)  K_{n-1}\left(  \tau_{n}\right)  . \label{2.9}%
\end{equation}

Repeatedly using Eq. (\ref{2.9}), we obtain%
\begin{align}
K_{n}\left(  \tau\right)   &  =\int_{0}^{\tau}d\tau_{n}\int_{0}^{\tau_{n}%
}d\tau_{n-1}\cdots\int_{0}^{\tau_{3}}d\tau_{2}\int_{0}^{\tau_{2}}d\tau
_{1}\nonumber\\
\times &  K_{0}\left(  \tau-\tau_{n}\right)  \left(  -V\right)  K_{0}\left(
\tau_{n}-\tau_{n-1}\right)  \left(  -V\right)  \cdots\left(  -V\right)
K_{0}\left(  \tau_{2}-\tau_{1}\right)  \left(  -V\right)  K_{0}\left(
\tau_{1}\right)  .\label{2.10}%
\end{align}

\section{The off-diagonal heat kernel \label{Matrix}}

Before we calculate the $n$-th order matrix elements of\ heat kernels, we
first need to calculate the matrix elements of the leading order heat kernel.

The matrix element of a leading order heat kernel in the coordinate
representation reads \cite{mukhanov2007introduction}%

\begin{equation}
\left\langle x\left\vert K_{0}\left(  \tau\right)  \right\vert y\right\rangle
=\frac{1}{\left(  4\pi\tau\right)  ^{\nu/2}}e^{-\left(  x-y\right)
^{2}/\left(  4\tau\right)  }, \label{3.4}%
\end{equation}
where $\nu$ denotes the dimension of space.

Now we calculate the $n$-th order matrix elements of a heat kernel
$\left\langle x\left\vert K_{n}\right\vert y\right\rangle $ based on Eq.
(\ref{2.10}).

Inserting the identity $1=\int d^{\nu}z\left\vert z\right\rangle \left\langle
z\right\vert $ into $\left\langle x\left\vert K_{n}\right\vert y\right\rangle
$, we have%

\begin{align}
\left\langle x\left\vert K_{n}\right\vert y\right\rangle  &  =\left(
-1\right)  ^{n}\int_{0}^{\tau}d\tau_{n}\int_{0}^{\tau_{n}}d\tau_{n-1}%
\cdots\int_{0}^{\tau_{3}}d\tau_{2}\int_{0}^{\tau_{2}}d\tau_{1}\int d^{\nu
}z_{2n}\int d^{\nu}z_{2n-1}\cdots\int d^{\nu}z_{2}\int d^{\nu}z_{1}\nonumber\\
&  \times\left\langle x\left\vert K_{0}\left(  \tau-\tau_{n}\right)
\right\vert z_{2n}\right\rangle \left\langle z_{2n}\left\vert V\right\vert
z_{2n-1}\right\rangle \left\langle z_{2n-1}\left\vert K_{0}\left(  \tau
_{n}-\tau_{n-1}\right)  \right\vert z_{2n-2}\right\rangle \cdots\left\langle
z_{2}\left\vert V\right\vert z_{1}\right\rangle \left\langle z_{1}\left\vert
K_{0}\left(  \tau_{1}\right)  \right\vert y\right\rangle .\label{6.6}%
\end{align}
Substituting $\left\langle x\left\vert V\right\vert y\right\rangle =V\left(
x\right)  \delta\left(  x-y\right)  $ into Eq. (\ref{6.6}) gives
\begin{align}
\left\langle x\left\vert K_{n}\right\vert y\right\rangle  &  =\left(
-1\right)  ^{n}\int_{0}^{\tau}d\tau_{n}\int_{0}^{\tau_{n}}d\tau_{n-1}%
\cdots\int_{0}^{\tau_{3}}d\tau_{2}\int_{0}^{\tau_{2}}d\tau_{1}\int d^{\nu
}z_{2n}\int d^{\nu}z_{2n-2}\cdots\int d^{\nu}z_{2}\nonumber\label{6.8}\\
&  \times\left\langle x\left\vert K_{0}\left(  \tau-\tau_{n}\right)
\right\vert z_{2n}\right\rangle V\left(  z_{2n}\right)  \left\langle
z_{2n}\left\vert K_{0}\left(  \tau_{n}-\tau_{n-1}\right)  \right\vert
z_{2n-2}\right\rangle V\left(  z_{2n-2}\right)  \cdots V\left(  z_{2}\right)
\left\langle z_{2}\left\vert K_{0}\left(  \tau_{1}\right)  \right\vert
y\right\rangle .
\end{align}
By Eq. (\ref{3.4}), we arrive at%
\begin{align}
\left\langle x\left\vert K_{n}\right\vert y\right\rangle  &  =\left(
-1\right)  ^{n}\int_{0}^{\tau}d\tau_{n}\int_{0}^{\tau_{n}}d\tau_{n-1}%
\cdots\int_{0}^{\tau_{3}}d\tau_{2}\int_{0}^{\tau_{2}}d\tau_{1}\int d^{\nu
}z_{2n}\int d^{\nu}z_{2n-2}\cdots\int d^{\nu}z_{2}\nonumber\label{6.11}\\
&  \times\frac{\exp\left(  -\frac{\left(  x-z_{2n}\right)  ^{2}}{4\left(
\tau-\tau_{n}\right)  }\right)  }{\left[  4\pi\left(  \tau-\tau_{n}\right)
\right]  ^{\nu/2}}V\left(  z_{2n}\right)  \frac{\exp\left(  -\frac{\left(
z_{2n}-z_{2n-2}\right)  ^{2}}{4\left(  \tau_{n}-\tau_{n-1}\right)  }\right)
}{\left[  4\pi\left(  \tau_{n}-\tau_{n-1}\right)  \right]  ^{\nu/2}}V\left(
z_{2n-2}\right)  \cdots V\left(  z_{2}\right)  \frac{\exp\left(
-\frac{\left(  z_{2}-y\right)  ^{2}}{4\tau_{1}}\right)  }{\left(  4\pi\tau
_{1}\right)  ^{\nu/2}}.
\end{align}
The Fourier transform of $V\left(  z\right)  $ is%
\begin{equation}
V\left(  z\right)  =\int\frac{d^{\nu}k}{\left(  2\pi\right)  ^{\nu/2}}%
e^{ikz}\tilde{V}\left(  k\right)  .\label{6.13}%
\end{equation}
Substituting Eq. (\ref{6.13}) into Eq. (\ref{6.11}), we obtain%
\begin{align}
\left\langle x\left\vert K_{n}\right\vert y\right\rangle  &  =\left(
-1\right)  ^{n}\int_{0}^{\tau}d\tau_{n}\int_{0}^{\tau_{n}}d\tau_{n-1}%
\cdots\int_{0}^{\tau_{3}}d\tau_{2}\int_{0}^{\tau_{2}}d\tau_{1}\nonumber\\
\times &  \int\frac{d^{\nu}k_{n}}{\left(  2\pi\right)  ^{\nu/2}}\tilde
{V}\left(  k_{n}\right)  \int\frac{d^{\nu}k_{n-1}}{\left(  2\pi\right)
^{\nu/2}}\tilde{V}\left(  k_{n-1}\right)  \cdots\int\frac{d^{\nu}k_{1}%
}{\left(  2\pi\right)  ^{\nu/2}}\tilde{V}\left(  k_{1}\right)  I_{1},
\end{align}
where%
\begin{align}
I_{1} &  =\int d^{\nu}z_{2n}\int d^{\nu}z_{2n-2}\cdots\int d^{\nu}%
z_{2}\nonumber\\
&  \frac{\exp\left(  -\frac{\left(  x-z_{2n}\right)  ^{2}}{4\left(  \tau
-\tau_{n}\right)  }-\frac{\left(  z_{2n}-z_{2n-2}\right)  ^{2}}{4\left(
\tau_{n}-\tau_{n-1}\right)  }-\cdots-\frac{\left(  z_{2}-y\right)  ^{2}}%
{4\tau_{1}}+ik_{n}z_{2n}+ik_{n}z_{2n-2}+\cdots+ik_{1}z_{2}\right)  }{\left[
4\pi\left(  \tau-\tau_{n}\right)  \right]  ^{\nu/2}\left[  4\pi\left(
\tau_{n}-\tau_{n-1}\right)  \right]  ^{\nu/2}\cdots\left(  4\pi\tau
_{1}\right)  ^{\nu/2}}.\label{6.15}%
\end{align}

The integral in $I_{1}$ can be performed by using the Gaussian integral
\cite{mukhanov2007introduction},
\begin{align}
&  \int d^{\nu}x\exp\left(  -A\left\vert x-a\right\vert ^{2}-B\left\vert
x-b\right\vert ^{2}+2cx\right) \nonumber\\
&  =\frac{\pi^{\nu/2}}{\left(  A+B\right)  ^{\nu/2}}\exp\left(  -\frac
{AB\left\vert a-b\right\vert ^{2}}{A+B}+\frac{2c\left(  Aa+Bb\right)
+\left\vert c\right\vert ^{2}}{A+B}\right)  . \label{6.16}%
\end{align}
Then we arrive at%
\begin{align}
\left\langle x\left\vert K_{n}\right\vert y\right\rangle  &  =\left(
-1\right)  ^{n}\frac{e^{-\left(  x-y\right)  ^{2}/\left(  4\tau\right)  }%
}{\left(  4\pi\tau\right)  ^{\nu/2}}\int_{0}^{\tau}d\tau_{n}\cdots\int%
_{0}^{\tau_{3}}d\tau_{2}\int_{0}^{\tau_{2}}d\tau_{1}\int\frac{d^{\nu}k_{n}%
}{\left(  2\pi\right)  ^{\nu/2}}\tilde{V}\left(  k_{n}\right)  \cdots\int%
\frac{d^{\nu}k_{1}}{\left(  2\pi\right)  ^{\nu/2}}\tilde{V}\left(
k_{1}\right) \nonumber\label{6.17}\\
&  \times\exp\left(  ik_{1}\left(  \frac{\tau_{1}}{\tau}x+\frac{\tau-\tau_{1}%
}{\tau}y\right)  +ik_{2}\left(  \frac{\tau_{2}}{\tau}x+\frac{\tau-\tau_{2}%
}{\tau}y\right)  +\cdots\right. \nonumber\\
&  +ik_{n-1}\left(  \frac{\tau_{n-1}}{\tau}x+\frac{\tau-\tau_{n-1}}{\tau
}y\right)  +ik_{n}\left(  \frac{\tau_{n}}{\tau}x+\frac{\tau-\tau_{n}}{\tau
}y\right) \nonumber\\
&  -\left[  \frac{\left(  \tau-\tau_{1}\right)  \tau_{1}}{\tau}k_{1}^{2}%
+\frac{\left(  \tau-\tau_{2}\right)  \tau_{2}}{\tau}k_{2}^{2}+\cdots
+\frac{\left(  \tau-\tau_{n-1}\right)  \tau_{n-1}}{\tau}k_{n-1}^{2}%
+\frac{\left(  \tau-\tau_{n}\right)  \tau_{n}}{\tau}k_{n}^{2}\right]
\nonumber\\
&  -2\left[  \frac{\left(  \tau-\tau_{2}\right)  \tau_{1}}{\tau}k_{1}%
k_{2}+\cdots+\frac{\left(  \tau-\tau_{n-1}\right)  \tau_{1}}{\tau}k_{1}%
k_{n-1}+\frac{\left(  \tau-\tau_{n}\right)  \tau_{1}}{\tau}k_{1}k_{n}\right]
\nonumber\\
&  -2\left[  \frac{\left(  \tau-\tau_{3}\right)  \tau_{2}}{\tau}k_{2}%
k_{3}+\cdots+\frac{\left(  \tau-\tau_{n-1}\right)  \tau_{2}}{\tau}k_{2}%
k_{n-1}+\frac{\left(  \tau-\tau_{n}\right)  \tau_{2}}{\tau}k_{2}k_{n}\right]
-\cdots\nonumber\\
&  -2\left[  \frac{\left(  \tau-\tau_{n-1}\right)  \tau_{n-2}}{\tau}%
k_{n-2}k_{n-1}+\frac{\left(  \tau-\tau_{n}\right)  \tau_{n-2}}{\tau}%
k_{n-2}k_{n}\right]  \left.  -2\frac{\left(  \tau-\tau_{n}\right)  \tau_{n-1}%
}{\tau}k_{n-1}k_{n}\right)  .
\end{align}
Introducing%

\begin{equation}
R_{1}=\frac{\tau_{1}}{\tau}x+\frac{\tau-\tau_{1}}{\tau}y,R_{2}=\frac{\tau_{2}%
}{\tau}x+\frac{\tau-\tau_{2}}{\tau}y,\cdots,R_{n}=\frac{\tau_{n}}{\tau}%
x+\frac{\tau-\tau_{n}}{\tau}y \label{6.18}%
\end{equation}
and using Eq. (\ref{6.13}), we obtain%

\begin{align}
\left\langle x\left\vert K_{n}\right\vert y\right\rangle  &  =\left(
-1\right)  ^{n}\frac{e^{-\left(  x-y\right)  ^{2}/\left(  4\tau\right)  }%
}{\left(  4\pi\tau\right)  ^{\nu/2}}\int_{0}^{\tau}d\tau_{n}\cdots\int%
_{0}^{\tau_{3}}d\tau_{2}\int_{0}^{\tau_{2}}d\tau_{1}\nonumber\label{6.19}\\
&  \times\exp\left(  \frac{\left(  \tau-\tau_{1}\right)  \tau_{1}}{\tau}%
\nabla_{R_{1}}^{2}+\cdots+\frac{\left(  \tau-\tau_{n-1}\right)  \tau_{n-1}%
}{\tau}\nabla_{R_{n-1}}^{2}+\frac{\left(  \tau-\tau_{n}\right)  \tau_{n}}%
{\tau}\nabla_{R_{n}}^{2}\right. \nonumber\\
&  +2\left[  \frac{\left(  \tau-\tau_{2}\right)  \tau_{1}}{\tau}\nabla_{R_{1}%
}\nabla_{R_{2}}+\cdots+\frac{\left(  \tau-\tau_{n-1}\right)  \tau_{1}}{\tau
}\nabla_{R_{1}}\nabla_{R_{n-1}}+\frac{\left(  \tau-\tau_{n}\right)  \tau_{1}%
}{\tau}\nabla_{R_{1}}\nabla_{R_{n}}\right] \nonumber\\
&  +2\left[  \frac{\left(  \tau-\tau_{3}\right)  \tau_{2}}{\tau}\nabla_{R_{2}%
}\nabla_{R_{3}}+\cdots+\frac{\left(  \tau-\tau_{n-1}\right)  \tau_{2}}{\tau
}\nabla_{R_{2}}\nabla_{R_{n-1}}+\frac{\left(  \tau-\tau_{n}\right)  \tau_{2}%
}{\tau}\nabla_{R_{2}}\nabla_{R_{n}}\right] \nonumber\\
&  \left.  +\cdots+2\frac{\left(  \tau-\tau_{n}\right)  \tau_{n-1}}{\tau
}\nabla_{R_{n-1}}\nabla_{R_{n}}\right)  V\left(  R_{1}\right)  V\left(
R_{2}\right)  \cdots V\left(  R_{n}\right)  .
\end{align}

\section{An alternative expression of the $n$-th order heat kernels
\label{alternative}}

In this section, we suggest an alternative expression of the $n$-th order
matrix element of heat kernel $\left\langle x\left\vert K_{n}\right\vert
y\right\rangle $, in which the integral with variable upper limit given in the
above section is replaced by definite integrals.

The integral in Eq. (\ref{6.19}) is a variable upper limit integral. In the
following we rewrite Eq. (\ref{6.19}) by a definite integral.

Introducing%
\begin{equation}
\tau_{1}=u_{1}\tau_{2}\text{, \ }\tau_{2}=u_{2}\tau_{3}\text{, }\cdots
\tau_{n-1}=u_{n-1}\tau_{n}\text{, \ }\tau_{n}=u_{n}\tau, \label{6.34}%
\end{equation}
we can rewrite Eq. (\ref{6.19}) as%

\begin{align}
\left\langle x\left\vert K_{n}\right\vert y\right\rangle  &  =\left(
-1\right)  ^{n}\frac{e^{-\left(  x-y\right)  ^{2}/\left(  4\tau\right)  }%
}{\left(  4\pi\tau\right)  ^{\nu/2}}\tau^{n}\int_{0}^{1}u_{n}^{n-1}du_{n}%
\int_{0}^{1}u_{n-1}^{n-2}du_{n-1}\cdots\int_{0}^{1}u_{2}du_{2}\int_{0}%
^{1}du_{1}\nonumber\label{6.36}\\
&  \times\exp\left(  \left(  1-u_{1}u_{2}\cdots u_{n}\right)  u_{1}u_{2}\cdots
u_{n}\tau\nabla_{R_{1}}^{2}+\cdots+\left(  1-u_{n-1}u_{n}\right)  u_{n-1}%
u_{n}\tau\nabla_{R_{n-1}}^{2}\right. \nonumber\\
&  +\left(  1-u_{n}\right)  u_{n}\tau\nabla_{R_{n}}^{2}+2\left[  \left(
1-u_{2}\cdots u_{n}\right)  u_{1}u_{2}\cdots u_{n}\tau\nabla_{R_{1}}%
\nabla_{R_{2}}+\cdots\right. \nonumber\\
&  +\left(  1-u_{n-1}u_{n}\right)  u_{1}u_{2}\cdots u_{n}\tau\nabla_{R_{1}%
}\nabla_{R_{n-1}}\left.  +\left(  1-u_{n}\right)  u_{1}u_{2}\cdots u_{n}%
\tau\nabla_{R_{1}}\nabla_{R_{n}}\right] \nonumber\\
&  +2\left[  \left(  1-u_{3}\cdots u_{n}\right)  u_{2}\cdots u_{n}\tau
\nabla_{R_{2}}\nabla_{R_{3}}+\cdots+\left(  1-u_{n-1}u_{n}\right)  u_{2}\cdots
u_{n}\tau\nabla_{R_{2}}\nabla_{R_{n-1}}\right. \nonumber\\
&  +\left.  \left(  1-u_{n}\right)  u_{2}\cdots u_{n}\tau\nabla_{R_{2}}%
\nabla_{R_{n}}\right]  +\cdots\nonumber\\
&  +2\left[  \left(  1-u_{n-1}u_{n}\right)  u_{n-2}u_{n-1}u_{n}\tau
\nabla_{R_{n-2}}\nabla_{R_{n-1}}+\left(  1-u_{n}\right)  u_{n-2}u_{n-1}%
u_{n}\tau\nabla_{R_{n-2}}\nabla_{R_{n}}\right] \nonumber\\
&  \left.  +2\left(  1-u_{n}\right)  u_{n-1}u_{n}\tau\nabla_{R_{n-1}}%
\nabla_{R_{n}}\right)  V\left(  R_{1}\right)  \cdots V\left(  R_{n-1}\right)
V\left(  R_{n}\right)  .
\end{align}

This expression of $\left\langle x\left\vert K_{n}\right\vert y\right\rangle $
also can be equivalently represented as%
\begin{align}
\left\langle x\left\vert K_{n}\right\vert y\right\rangle  &  =\left(
-1\right)  ^{n}\frac{e^{-\left(  x-y\right)  ^{2}/\left(  4\tau\right)  }%
}{\left(  4\pi\tau\right)  ^{\nu/2}}\tau^{n}\int_{0}^{1}u_{n}^{n-1}du_{n}%
\int_{0}^{1}u_{n-1}^{n-2}du_{n-1}\cdots\int_{0}^{1}u_{2}du_{2}\int_{0}%
^{1}du_{1}\nonumber\label{6.37}\\
&  \times\exp\left(  \left(
\begin{array}
[c]{ccccc}%
\left(  1-u_{1}u_{2}\cdots u_{n}\right)  \nabla_{R_{1}}, & \left(
1-u_{2}\cdots u_{n}\right)  \nabla_{R_{2}}, & \cdots, & \left(  1-u_{n-1}%
u_{n}\right)  \nabla_{R_{n-1}}, & \left(  1-u_{n}\right)  \nabla_{R_{n}}%
\end{array}
\right)  \right. \nonumber\\
&  \times\left.  \left(
\begin{array}
[c]{ccccc}%
1 & 0 & \cdots & 0 & 0\\
2 & 1 & \cdots & 0 & 0\\
\vdots & \vdots & \ddots & \vdots & \vdots\\
2 & 2 & \cdots & 1 & 0\\
2 & 2 & \cdots & 2 & 1
\end{array}
\right)  \left(
\begin{array}
[c]{c}%
u_{1}u_{2}\cdots u_{n}\tau\nabla_{R_{1}}\\
u_{2}\cdots u_{n}\tau\nabla_{R_{2}}\\
\vdots\\
u_{n-1}u_{n}\tau\nabla_{R_{n-1}}\\
u_{n}\tau\nabla_{R_{n}}%
\end{array}
\right)  \right)  V\left(  R_{1}\right)  \cdots V\left(  R_{n-1}\right)
V\left(  R_{n}\right)  .
\end{align}

\section{The first three orders of diagonal and global heat kernels \label{3}}

In this section, based on the above results, we give two equivalent represents
of the first three orders of diagonal and global heat kernels: one is
represented by $\tilde{V}\left(  k_{n}\right)  $ and the other is represented
by $V\left(  R_{n}\right)  $.

\subsection{The heat kernel with respect to $\tilde{V}\left(  k_{n}\right)  $}

\textit{First-order contribution.} The first-order heat kernel $\left\langle
x\left\vert K_{1}\right\vert y\right\rangle $ in the coordinate
representation, by Eq. (\ref{6.17}), is%
\begin{align}
\left\langle x\left\vert K_{1}\right\vert y\right\rangle  &  =-\frac
{e^{-\left(  x-y\right)  ^{2}/\left(  4\tau\right)  }}{\left(  4\pi
\tau\right)  ^{\nu/2}}\int_{0}^{\tau}d\tau_{1}\int\frac{d^{\nu}k}{\left(
2\pi\right)  ^{\nu/2}}\tilde{V}\left(  k\right) \nonumber\label{7.2}\\
&  \times\exp\left(  ik\frac{\tau_{1}x+\left(  \tau-\tau_{1}\right)  y}{\tau
}-\frac{\left(  \tau-\tau_{1}\right)  \tau_{1}}{\tau}k^{2}\right)  .
\end{align}

The first-order diagonal heat kernel is then%
\begin{equation}
\left\langle x\left\vert K_{1}\right\vert x\right\rangle =-\frac{1}{\left(
4\pi\tau\right)  ^{\nu/2}}\int_{0}^{\tau}d\tau_{1}\int\frac{d^{\nu}k}{\left(
2\pi\right)  ^{\nu/2}}\tilde{V}\left(  k\right)  \exp\left(  ikx-\frac{\left(
\tau-\tau_{1}\right)  \tau_{1}}{\tau}k^{2}\right)  . \label{7.21}%
\end{equation}

The first-order global heat kernel then reads
\begin{align}
K_{1}\left(  \tau\right)   &  =\int\left\langle x\left\vert K_{1}\right\vert
x\right\rangle d^{\nu}x\nonumber\label{7.22}\\
&  =-\frac{2\pi}{\left(  4\pi\tau\right)  ^{\nu/2}}\int_{0}^{\tau}d\tau
_{1}\int\frac{d^{\nu}k}{\left(  2\pi\right)  ^{\nu/2}}\tilde{V}\left(
k\right)  \exp\left(  -\frac{\left(  \tau-\tau_{1}\right)  \tau_{1}}{\tau
}k^{2}\right)  \delta\left(  k\right) \nonumber\\
&  =-\frac{1}{2^{\left(  3\nu-2\right)  /2}\pi^{\nu-1}\tau^{\left(
\nu-2\right)  /2}}\tilde{V}\left(  0\right)  .
\end{align}

\textit{Second-order contribution.} The second-order heat kernel $\left\langle
x\left\vert K_{2}\right\vert y\right\rangle $ in the coordinate
representation, by Eq. (\ref{6.17}), is%
\begin{align}
\left\langle x\left\vert K_{2}\right\vert y\right\rangle  &  =\frac
{e^{-\left(  x-y\right)  ^{2}/4\tau}}{\left(  4\pi\tau\right)  ^{\nu/2}}%
\int_{0}^{\tau}d\tau_{2}\int\frac{d^{\nu}k_{2}}{\left(  2\pi\right)  ^{\nu/2}%
}\tilde{V}\left(  k_{2}\right)  \int_{0}^{\tau_{2}}d\tau_{1}\int\frac{d^{\nu
}k_{1}}{\left(  2\pi\right)  ^{\nu/2}}\tilde{V}\left(  k_{1}\right)
\nonumber\label{7.3}\\
&  \times\exp\left(  ik_{1}\left(  \frac{\tau-\tau_{1}}{\tau}y+\frac{\tau_{1}%
}{\tau}x\right)  +ik_{2}\left(  \frac{\tau-\tau_{2}}{\tau}y+\frac{\tau_{2}%
}{\tau}x\right)  \right. \nonumber\\
&  \left.  -\frac{\left(  \tau-\tau_{1}\right)  \tau_{1}}{\tau}k_{1}^{2}%
-\frac{\left(  \tau-\tau_{2}\right)  \tau_{2}}{\tau}k_{2}^{2}-\frac{\left(
\tau-\tau_{2}\right)  \tau_{1}}{\tau}2k_{2}k_{1}\right)  .
\end{align}

The second-order diagonal heat kernel is%
\begin{align}
\left\langle x\left\vert K_{2}\right\vert x\right\rangle  &  =\frac{1}{\left(
4\pi\tau\right)  ^{\nu/2}}\int_{0}^{\tau}d\tau_{2}\int\frac{d^{\nu}k_{2}%
}{\left(  2\pi\right)  ^{\nu/2}}\tilde{V}\left(  k_{2}\right)  \int_{0}%
^{\tau_{2}}d\tau_{1}\int\frac{d^{\nu}k_{1}}{\left(  2\pi\right)  ^{\nu/2}%
}\tilde{V}\left(  k_{1}\right) \nonumber\label{7.31}\\
&  \times\exp\left(  ik_{1}x+ik_{2}x-\frac{\left(  \tau-\tau_{1}\right)
\tau_{1}}{\tau}k_{1}^{2}-\frac{\left(  \tau-\tau_{2}\right)  \tau_{2}}{\tau
}k_{2}^{2}-\frac{\left(  \tau-\tau_{2}\right)  \tau_{1}}{\tau}2k_{2}%
k_{1}\right)  .
\end{align}

The second-order global heat kernel then reads
\begin{align}
K_{2}\left(  \tau\right)   &  =\int\left\langle x\left\vert K_{2}\right\vert
x\right\rangle d^{\nu}x\nonumber\label{7.32}\\
&  =\frac{1}{\left(  4\pi\tau\right)  ^{\nu/2}}\int_{0}^{\tau}d\tau_{2}%
\int\frac{d^{\nu}k_{2}}{\left(  2\pi\right)  ^{\nu/2}}\tilde{V}\left(
k_{2}\right)  \int_{0}^{\tau_{2}}d\tau_{1}\int\frac{d^{\nu}k_{1}}{\left(
2\pi\right)  ^{\nu/2}}\tilde{V}\left(  k_{1}\right) \nonumber\\
&  \times\exp\left(  -\frac{\left(  \tau-\tau_{1}\right)  \tau_{1}}{\tau}%
k_{1}^{2}-\frac{\left(  \tau-\tau_{2}\right)  \tau_{2}}{\tau}k_{2}^{2}%
-\frac{\left(  \tau-\tau_{2}\right)  \tau_{1}}{\tau}2k_{2}k_{1}\right)
\delta\left(  k_{1}+k_{2}\right) \nonumber\\
&  =\frac{1}{2^{\left(  3\nu+2\right)  /2}\pi^{\left(  2\nu-1\right)  /2}%
\tau^{\left(  \nu-3\right)  /2}}\int\frac{d^{\nu}k_{2}}{\left(  2\pi\right)
^{\nu/2}}\frac{e^{-k_{2}^{2}\tau/4}}{k_{2}}\operatorname{erfi}\left(
\frac{k_{2}\sqrt{\tau}}{2}\right)  \tilde{V}^{2}\left(  k_{2}\right)  ,
\end{align}
where $\operatorname{erfi}\left(  z\right)  $ gives the imaginary error
function $\operatorname{erf}\left(  z\right)  /i$.

\textit{Third-order contribution. }The third-order heat kernel $\left\langle
x\left\vert K_{3}\right\vert y\right\rangle $ in the coordinate
representation, by Eq. (\ref{6.17}), is%
\begin{align}
\left\langle x\left\vert K_{3}\right\vert y\right\rangle  &  =-\frac
{e^{-\left(  x-y\right)  ^{2}/\left(  4\tau\right)  }}{\left(  4\pi
\tau\right)  ^{\nu/2}}\int_{0}^{\tau}d\tau_{3}\int\frac{d^{\nu}k_{3}}{\left(
2\pi\right)  ^{\nu/2}}\tilde{V}\left(  k_{3}\right)  \int_{0}^{\tau_{3}}%
d\tau_{2}\int\frac{d^{\nu}k_{2}}{\left(  2\pi\right)  ^{\nu/2}}\tilde
{V}\left(  k_{2}\right)  \int_{0}^{\tau_{2}}d\tau_{1}\int\frac{d^{\nu}k_{1}%
}{\left(  2\pi\right)  ^{\nu/2}}\tilde{V}\left(  k_{1}\right)
\nonumber\label{7.4}\\
&  \times\exp\left(  ik_{1}\left[  \frac{\tau_{1}}{\tau}x+\frac{\left(
\tau-\tau_{1}\right)  }{\tau}y\right]  +ik_{2}\left[  \frac{\tau_{2}}{\tau
}x+\frac{\left(  \tau-\tau_{2}\right)  }{\tau}y\right]  +ik_{3}\left[
\frac{\tau_{3}}{\tau}x+\frac{\left(  \tau-\tau_{3}\right)  }{\tau}y\right]
\right. \nonumber\\
&  -\left[  k_{1}^{2}\frac{\left(  \tau-\tau_{1}\right)  \tau_{1}}{\tau}%
+k_{2}^{2}\frac{\left(  \tau-\tau_{2}\right)  \tau_{2}}{\tau}+k_{3}^{2}%
\frac{\left(  \tau-\tau_{3}\right)  \tau_{3}}{\tau}\right] \nonumber\\
&  \left.  -\left[  2k_{1}k_{2}\frac{\left(  \tau-\tau_{2}\right)  \tau_{1}%
}{\tau}+2k_{1}k_{3}\frac{\left(  \tau-\tau_{3}\right)  \tau_{1}}{\tau}%
+2k_{2}k_{3}\frac{\left(  \tau-\tau_{3}\right)  \tau_{2}}{\tau}\right]
\right)  .
\end{align}

The third-order diagonal heat kernel $\left\langle x\left\vert K_{3}%
\right\vert x\right\rangle $ is%
\begin{align}
\left\langle x\left\vert K_{3}\right\vert x\right\rangle  &  =-\frac
{1}{\left(  4\pi\tau\right)  ^{\nu/2}}\int_{0}^{\tau}d\tau_{3}\int\frac
{d^{\nu}k_{3}}{\left(  2\pi\right)  ^{\nu/2}}\tilde{V}\left(  k_{3}\right)
\int_{0}^{\tau_{3}}d\tau_{2}\int\frac{d^{\nu}k_{2}}{\left(  2\pi\right)
^{\nu/2}}\tilde{V}\left(  k_{2}\right)  \int_{0}^{\tau_{2}}d\tau_{1}\int%
\frac{d^{\nu}k_{1}}{\left(  2\pi\right)  ^{\nu/2}}\tilde{V}\left(
k_{1}\right) \nonumber\label{7.41}\\
&  \times\exp\left(  ik_{1}x+ik_{2}x+ik_{3}x-\left[  k_{1}^{2}\frac{\left(
\tau-\tau_{1}\right)  \tau_{1}}{\tau}+k_{2}^{2}\frac{\left(  \tau-\tau
_{2}\right)  \tau_{2}}{\tau}+k_{3}^{2}\frac{\left(  \tau-\tau_{3}\right)
\tau_{3}}{\tau}\right]  \right. \nonumber\\
&  -\left.  \left[  2k_{1}k_{2}\frac{\left(  \tau-\tau_{2}\right)  \tau_{1}%
}{\tau}+2k_{1}k_{3}\frac{\left(  \tau-\tau_{3}\right)  \tau_{1}}{\tau}%
+2k_{2}k_{3}\frac{\left(  \tau-\tau_{3}\right)  \tau_{2}}{\tau}\right]
\right)  .
\end{align}

The third-order global heat kernel then reads
\begin{align}
K_{3}\left(  \tau\right)   &  =\int\left\langle x\left\vert K_{3}\right\vert
x\right\rangle d^{\nu}x\nonumber\label{7.42}\\
&  =-\frac{1}{\left(  4\pi\tau\right)  ^{\nu/2}}\int_{0}^{\tau}d\tau_{3}%
\int\frac{d^{\nu}k_{3}}{\left(  2\pi\right)  ^{\nu/2}}\tilde{V}\left(
k_{3}\right)  \int_{0}^{\tau_{3}}d\tau_{2}\int\frac{d^{\nu}k_{2}}{\left(
2\pi\right)  ^{\nu/2}}\tilde{V}\left(  k_{2}\right)  \int_{0}^{\tau_{2}}%
d\tau_{1}\int\frac{d^{\nu}k_{1}}{\left(  2\pi\right)  ^{\nu/2}}\tilde
{V}\left(  k_{1}\right) \nonumber\\
&  \times\exp\left(  -\left[  k_{1}^{2}\frac{\left(  \tau-\tau_{1}\right)
\tau_{1}}{\tau}+k_{2}^{2}\frac{\left(  \tau-\tau_{2}\right)  \tau_{2}}{\tau
}+k_{3}^{2}\frac{\tau_{3}\left(  \tau-\tau_{3}\right)  }{\tau}\right]  \right.
\nonumber\\
&  -\left.  \left[  2k_{1}k_{2}\frac{\left(  \tau-\tau_{2}\right)  \tau_{1}%
}{\tau}+2k_{1}k_{3}\frac{\left(  \tau-\tau_{3}\right)  \tau_{1}}{\tau}%
+2k_{2}k_{3}\frac{\left(  \tau-\tau_{3}\right)  \tau_{2}}{\tau}\right]
\right)  \delta\left(  k_{1}+k_{2}+k_{3}\right) \nonumber\\
&  =-\frac{1}{\left(  4\pi\tau\right)  ^{\nu/2}}\int_{0}^{\tau}d\tau_{3}%
\int\frac{d^{\nu}k_{3}}{\left(  2\pi\right)  ^{\nu/2}}\tilde{V}\left(
k_{3}\right)  \int_{0}^{\tau_{3}}d\tau_{2}\int\frac{d^{\nu}k_{2}}{\left(
2\pi\right)  ^{\nu/2}}\tilde{V}\left(  k_{2}\right)  \int_{0}^{\tau_{2}}%
d\tau_{1}\frac{1}{\left(  2\pi\right)  ^{\nu/2}}\tilde{V}\left(  -k_{2}%
-k_{3}\right) \nonumber\\
&  \times\exp\left(  \frac{\left(  \tau_{1}-\tau_{2}\right)  \left(  \tau
_{1}-\tau_{2}+\tau\right)  }{\tau}k_{2}^{2}+\frac{\left(  \tau_{1}-\tau
_{3}\right)  \left(  \tau_{1}-\tau_{3}+\tau\right)  }{\tau}k_{3}^{2}%
+\frac{\left(  \tau_{1}-\tau_{2}\right)  \left(  \tau_{1}-\tau_{3}%
+\tau\right)  }{\tau}2k_{2}k_{3}\right)  .
\end{align}

\subsection{The heat kernel with respect to $V\left(  R_{n}\right)  $}

In this section we give an alternative expression of first three orders of
diagonal and global heat kernels, which is represented $V\left(  R_{n}\right)
$, based on Eq. (\ref{6.19}).

By Eq. (\ref{6.19}), the first-order heat kernel $\left\langle x\left\vert
K_{1}\right\vert y\right\rangle $ is%

\begin{equation}
\left\langle x\left\vert K_{1}\right\vert y\right\rangle =-\frac{e^{-\left(
x-y\right)  ^{2}/\left(  4\tau\right)  }}{\left(  4\pi\tau\right)  ^{\nu/2}%
}\int_{0}^{\tau}d\tau_{1}\exp\left(  \frac{\left(  \tau-\tau_{1}\right)
\tau_{1}}{\tau}\nabla_{R_{1}}^{2}\right)  V\left(  R_{1}\right)  .\label{7.6}%
\end{equation}
The first-order diagonal heat kernel is%

\begin{equation}
\left\langle x\left\vert K_{1}\right\vert x\right\rangle =-\frac{1}{\left(
4\pi\tau\right)  ^{\nu/2}}\int_{0}^{\tau}d\tau_{1}\exp\left(  \frac{\left(
\tau-\tau_{1}\right)  \tau_{1}}{\tau}\nabla_{x}^{2}\right)  V\left(  x\right)
.
\end{equation}
The first-order global heat kernel is
\begin{align}
K_{1}\left(  \tau\right)   &  =\int d^{\nu}x\left\langle x\left\vert
K_{1}\right\vert x\right\rangle \nonumber\\
&  =-\frac{1}{\left(  4\pi\tau\right)  ^{\nu/2}}\int_{0}^{\tau}d\tau_{1}\int
d^{\nu}x\exp\left(  \frac{\left(  \tau-\tau_{1}\right)  \tau_{1}}{\tau}%
\nabla_{x}^{2}\right)  V\left(  x\right)  .
\end{align}
The second-order heat kernel is%
\begin{align}
\left\langle x\left\vert K_{2}\right\vert y\right\rangle  &  =\frac
{e^{-\left(  x-y\right)  ^{2}/\left(  4\tau\right)  }}{\left(  4\pi
\tau\right)  ^{\nu/2}}\int_{0}^{\tau}d\tau_{2}\int_{0}^{\tau_{2}}d\tau_{1}%
\exp\left(  \frac{\left(  \tau-\tau_{1}\right)  \tau_{1}}{\tau}\nabla_{R_{1}%
}^{2}\right.  \nonumber\label{7.7}\\
&  +\frac{\left(  \tau-\tau_{2}\right)  \tau_{2}}{\tau}\nabla_{R_{2}}%
^{2}+\left.  2\frac{\left(  \tau-\tau_{2}\right)  \tau_{1}}{\tau}\nabla
_{R_{1}}\nabla_{R_{2}}\right)  V\left(  R_{2}\right)  V\left(  R_{1}\right)  .
\end{align}
The second-order diagonal heat kernel is
\begin{align}
\left\langle x\left\vert K_{2}\right\vert x\right\rangle  &  =\frac{1}{\left(
4\pi\tau\right)  ^{\nu/2}}\int_{0}^{\tau}d\tau_{2}\int_{0}^{\tau_{2}}d\tau
_{1}\exp\left(  \frac{\left(  \tau-\tau_{1}\right)  \tau_{1}}{\tau}\nabla
_{x}^{2}\right.  \nonumber\\
&  +\frac{\left(  \tau-\tau_{2}\right)  \tau_{2}}{\tau}\nabla_{x}^{2}+\left.
2\frac{\left(  \tau-\tau_{2}\right)  \tau_{1}}{\tau}\nabla_{x}^{2}\right)
V^{2}\left(  x\right)  \nonumber\\
&  =\frac{1}{\left(  4\pi\tau\right)  ^{\nu/2}}\int_{0}^{\tau}d\tau_{2}%
\int_{0}^{\tau_{2}}d\tau_{1}\exp\left(  \frac{\tau\left(  3\tau_{1}+\tau
_{2}\right)  -\left(  \tau_{1}+\tau_{2}\right)  ^{2}}{\tau}\nabla_{x}%
^{2}\right)  V^{2}\left(  x\right)  .
\end{align}
The second-order global heat kernel is
\begin{align}
K_{2}\left(  \tau\right)   &  =\int d^{\nu}x\left\langle x\left\vert
K_{2}\right\vert x\right\rangle \nonumber\\
&  =\frac{1}{\left(  4\pi\tau\right)  ^{\nu/2}}\int_{0}^{\tau}d\tau_{2}%
\int_{0}^{\tau_{2}}d\tau_{1}\int d^{\nu}x\exp\left(  \frac{\left(  \tau
-\tau_{1}\right)  \tau_{1}}{\tau}\nabla_{x}^{2}\right.  \nonumber\\
&  +\frac{\left(  \tau-\tau_{2}\right)  \tau_{2}}{\tau}\nabla_{x}^{2}+\left.
2\frac{\left(  \tau-\tau_{2}\right)  \tau_{1}}{\tau}\nabla_{x}^{2}\right)
V^{2}\left(  x\right)  \nonumber\\
&  =\frac{1}{\left(  4\pi\tau\right)  ^{\nu/2}}\int_{0}^{\tau}d\tau_{2}%
\int_{0}^{\tau_{2}}d\tau_{1}\int d^{\nu}x\exp\left(  \frac{\tau\left(
3\tau_{1}+\tau_{2}\right)  -\left(  \tau_{1}+\tau_{2}\right)  ^{2}}{\tau
}\nabla_{x}^{2}\right)  V^{2}\left(  x\right)  .
\end{align}
The third-order heat kernel $\left\langle x\left\vert K_{3}\right\vert
y\right\rangle $ is%

\begin{align}
\left\langle x\left\vert K_{3}\right\vert y\right\rangle  &  =-\frac
{e^{-\left(  x-y\right)  ^{2}/\left(  4\tau\right)  }}{\left(  4\pi
\tau\right)  ^{\nu/2}}\int_{0}^{\tau}d\tau_{3}\int_{0}^{\tau_{3}}d\tau_{2}%
\int_{0}^{\tau_{2}}d\tau_{1}\nonumber\label{7.8}\\
&  \times\exp\left(  \frac{\left(  \tau-\tau_{1}\right)  \tau_{1}}{\tau}%
\nabla_{R_{1}}^{2}+\frac{\left(  \tau-\tau_{2}\right)  \tau_{2}}{\tau}%
\nabla_{R_{2}}^{2}+\frac{\left(  \tau-\tau_{3}\right)  \tau_{3}}{\tau}%
\nabla_{R_{3}}^{2}+2\frac{\left(  \tau-\tau_{2}\right)  \tau_{1}}{\tau}%
\nabla_{R_{1}}\nabla_{R_{2}}\right.  \nonumber\\
&  +\left.  2\frac{\left(  \tau-\tau_{3}\right)  \tau_{1}}{\tau}\nabla_{R_{1}%
}\nabla_{R_{3}}+2\frac{\left(  \tau-\tau_{3}\right)  \tau_{2}}{\tau}%
\nabla_{R_{2}}\nabla_{R_{3}}\right)  V\left(  R_{1}\right)  V\left(
R_{2}\right)  V\left(  R_{3}\right)  .
\end{align}
The third-order diagonal heat kernel is
\begin{align}
\left\langle x\left\vert K_{3}\right\vert x\right\rangle  &  =-\frac
{1}{\left(  4\pi\tau\right)  ^{\nu/2}}\int_{0}^{\tau}d\tau_{3}\int_{0}%
^{\tau_{3}}d\tau_{2}\int_{0}^{\tau_{2}}d\tau_{1}\nonumber\\
&  \times\exp\left(  \frac{\left(  \tau-\tau_{1}\right)  \tau_{1}}{\tau}%
\nabla_{x}^{2}+\frac{\left(  \tau-\tau_{2}\right)  \tau_{2}}{\tau}\nabla
_{x}^{2}+\frac{\left(  \tau-\tau_{3}\right)  \tau_{3}}{\tau}\nabla_{x}%
^{2}+2\frac{\left(  \tau-\tau_{2}\right)  \tau_{1}}{\tau}\nabla_{x}%
^{2}\right.  \nonumber\\
&  +\left.  2\frac{\left(  \tau-\tau_{3}\right)  \tau_{1}}{\tau}\nabla_{x}%
^{2}+2\frac{\left(  \tau-\tau_{3}\right)  \tau_{2}}{\tau}\nabla_{x}%
^{2}\right)  V^{3}\left(  x\right)  \nonumber\\
&  =-\frac{1}{\left(  4\pi\tau\right)  ^{\nu/2}}\int_{0}^{\tau}d\tau_{3}%
\int_{0}^{\tau_{3}}d\tau_{2}\int_{0}^{\tau_{2}}d\tau_{1}\nonumber\\
&  \times\exp\left(  \frac{\tau\left(  5\tau_{1}+3\tau_{2}+\tau_{3}\right)
-\left(  \tau_{1}+\tau_{2}+\tau_{3}\right)  ^{2}}{\tau}\nabla_{x}^{2}\right)
V^{3}\left(  x\right)  .
\end{align}
The third-order global heat kernel is%
\begin{align}
K_{3}\left(  \tau\right)   &  =\int\left\langle x\left\vert K_{3}\right\vert
x\right\rangle d^{\nu}x\nonumber\\
&  =-\frac{1}{\left(  4\pi\tau\right)  ^{\nu/2}}\int_{0}^{\tau}d\tau_{3}%
\int_{0}^{\tau_{3}}d\tau_{2}\int_{0}^{\tau_{2}}d\tau_{1}\int d^{\nu
}x\nonumber\\
&  \times\exp\left(  \frac{\left(  \tau-\tau_{1}\right)  \tau_{1}}{\tau}%
\nabla_{x}^{2}+\frac{\left(  \tau-\tau_{2}\right)  \tau_{2}}{\tau}\nabla
_{x}^{2}+\frac{\left(  \tau-\tau_{3}\right)  \tau_{3}}{\tau}\nabla_{x}%
^{2}+2\frac{\left(  \tau-\tau_{2}\right)  \tau_{1}}{\tau}\nabla_{x}%
^{2}\right.  \nonumber\\
&  +\left.  2\frac{\left(  \tau-\tau_{3}\right)  \tau_{1}}{\tau}\nabla_{x}%
^{2}+2\frac{\left(  \tau-\tau_{3}\right)  \tau_{2}}{\tau}\nabla_{x}%
^{2}\right)  V^{3}\left(  x\right)  .\nonumber\\
&  =-\frac{1}{\left(  4\pi\tau\right)  ^{\nu/2}}\int_{0}^{\tau}d\tau_{3}%
\int_{0}^{\tau_{3}}d\tau_{2}\int_{0}^{\tau_{2}}d\tau_{1}\int d^{\nu
}x\nonumber\\
&  \times\exp\left(  \frac{\tau\left(  5\tau_{1}+3\tau_{2}+\tau_{3}\right)
-\left(  \tau_{1}+\tau_{2}+\tau_{3}\right)  ^{2}}{\tau}\nabla_{x}^{2}\right)
V^{3}\left(  x\right)  .
\end{align}

\subsection{The heat kernel with respect to $u_{n}$}

In this section, we give an equivalent expression of the first three orders of
diagonal and global heat kernels, which is represented $u_{n}$, based on Eq.
(\ref{6.36}).

By Eq. (\ref{6.36}), the first-order heat kernel $\left\langle x\left\vert
K_{1}\right\vert y\right\rangle $ is%
\begin{equation}
\left\langle x\left\vert K_{1}\right\vert y\right\rangle =-\frac{e^{-\left(
x-y\right)  ^{2}/\left(  4\tau\right)  }}{\left(  4\pi\tau\right)  ^{\nu/2}%
}\tau\int_{0}^{1}du_{1}\exp\left(  \left(  1-u_{1}\right)  u_{1}\tau
\nabla_{R_{1}}^{2}\right)  V\left(  R_{1}\right)  .\label{7.10}%
\end{equation}
The first-order diagonal heat kernel is
\begin{equation}
\left\langle x\left\vert K_{1}\right\vert x\right\rangle =-\frac{1}{\left(
4\pi\tau\right)  ^{\nu/2}}\tau\int_{0}^{1}du_{1}\exp\left(  \left(
1-u_{1}\right)  u_{1}\tau\nabla_{x}^{2}\right)  V\left(  x\right)  .
\end{equation}
The first-order global heat kernel is%
\begin{align}
K_{1}\left(  \tau\right)   &  =\int\left\langle x\left\vert K_{1}\right\vert
x\right\rangle d^{\nu}x\nonumber\\
&  =-\frac{1}{\left(  4\pi\tau\right)  ^{\nu/2}}\tau\int_{0}^{1}du_{1}\int
d^{\nu}x\exp\left(  \left(  1-u_{1}\right)  u_{1}\tau\nabla_{x}^{2}\right)
V\left(  x\right)  .
\end{align}
The second-order heat kernel $\left\langle x\left\vert K_{2}\right\vert
y\right\rangle $ is
\begin{align}
\left\langle x\left\vert K_{2}\right\vert y\right\rangle  &  =\frac
{e^{-\left(  x-y\right)  ^{2}/\left(  4\tau\right)  }}{\left(  4\pi
\tau\right)  ^{\nu/2}}\tau^{2}\int_{0}^{1}u_{2}du_{2}\int_{0}^{1}du_{1}%
\exp\left(  \left(  1-u_{1}u_{2}\right)  u_{1}u_{2}\tau\nabla_{R_{1}}%
^{2}\right.  \nonumber\label{7.11}\\
&  +\left(  1-u_{2}\right)  u_{2}\tau\nabla_{R_{2}}^{2}+\left.  2\left(
1-u_{2}\right)  u_{1}u_{2}\tau\nabla_{R_{1}}\nabla_{R_{2}}\right)  V\left(
R_{2}\right)  V\left(  R_{1}\right)  .
\end{align}
The second-order diagonal heat kernel is
\begin{align}
\left\langle x\left\vert K_{2}\right\vert x\right\rangle  &  =\frac{1}{\left(
4\pi\tau\right)  ^{\nu/2}}\tau^{2}\int_{0}^{1}u_{2}du_{2}\int_{0}^{1}%
du_{1}\exp\left(  \left(  1-u_{1}u_{2}\right)  u_{1}u_{2}\tau\nabla_{x}%
^{2}\right.  \nonumber\\
&  +\left(  1-u_{2}\right)  u_{2}\tau\nabla_{x}^{2}+\left.  2\left(
1-u_{2}\right)  u_{1}u_{2}\tau\nabla_{x}^{2}\right)  V^{2}\left(  x\right)  .
\end{align}
The second-order global heat kernel is
\begin{align}
K_{2}\left(  \tau\right)   &  =\int\left\langle x\left\vert K_{2}\right\vert
x\right\rangle d^{\nu}x\nonumber\\
&  =\frac{1}{\left(  4\pi\tau\right)  ^{\nu/2}}\tau^{2}\int_{0}^{1}u_{2}%
du_{2}\int_{0}^{1}du_{1}\int d^{\nu}x\exp\left(  \left(  1-u_{1}u_{2}\right)
u_{1}u_{2}\tau\nabla_{x}^{2}\right.  \nonumber\\
&  +\left(  1-u_{2}\right)  u_{2}\tau\nabla_{x}^{2}+\left.  2\left(
1-u_{2}\right)  u_{1}u_{2}\tau\nabla_{x}^{2}\right)  V^{2}\left(  x\right)  .
\end{align}
The third-order heat kernel $\left\langle x\left\vert K_{3}\right\vert
y\right\rangle $ is
\begin{align}
\left\langle x\left\vert K_{3}\right\vert y\right\rangle  &  =-\frac
{e^{-\left(  x-y\right)  ^{2}/\left(  4\tau\right)  }}{\left(  4\pi
\tau\right)  ^{\nu/2}}\tau^{3}\int_{0}^{1}u_{3}^{2}du_{3}\int_{0}^{1}%
u_{2}du_{2}\int_{0}^{1}du_{1}\nonumber\label{7.12}\\
&  \times\exp\left(  \left(  1-u_{1}u_{2}u_{3}\right)  u_{1}u_{2}u_{3}%
\tau\nabla_{R_{1}}^{2}+\left(  1-u_{2}u_{3}\right)  u_{2}u_{3}\tau
\nabla_{R_{2}}^{2}\right.  \nonumber\\
&  +\left(  1-u_{3}\right)  u_{3}\tau\nabla_{R_{3}}^{2}+2\left(  1-u_{2}%
u_{3}\right)  u_{1}u_{2}u_{3}\tau\nabla_{R_{1}}\nabla_{R_{2}}\nonumber\\
&  +\left.  2\left(  1-u_{3}\right)  u_{1}u_{2}u_{3}\tau\nabla_{R_{1}}%
\nabla_{R_{3}}+2\left(  1-u_{3}\right)  u_{2}u_{3}\tau\nabla_{R_{2}}%
\nabla_{R_{3}}\right)  V\left(  R_{1}\right)  V\left(  R_{2}\right)  V\left(
R_{3}\right)  .
\end{align}
The third-order diagonal heat kernel is
\begin{align}
\left\langle x\left\vert K_{3}\right\vert x\right\rangle  &  =-\frac
{1}{\left(  4\pi\tau\right)  ^{\nu/2}}\tau^{3}\int_{0}^{1}u_{3}^{2}du_{3}%
\int_{0}^{1}u_{2}du_{2}\int_{0}^{1}du_{1}\nonumber\\
&  \times\exp\left(  \left(  1-u_{1}u_{2}u_{3}\right)  u_{1}u_{2}u_{3}%
\tau\nabla_{x}^{2}+\left(  1-u_{2}u_{3}\right)  u_{2}u_{3}\tau\nabla_{x}%
^{2}\right.  \nonumber\\
&  +\left(  1-u_{3}\right)  u_{3}\tau\nabla_{x}^{2}+2\left(  1-u_{2}%
u_{3}\right)  u_{1}u_{2}u_{3}\tau\nabla_{x}^{2}\nonumber\\
&  +\left.  2\left(  1-u_{3}\right)  u_{1}u_{2}u_{3}\tau\nabla_{x}%
^{2}+2\left(  1-u_{3}\right)  u_{2}u_{3}\tau\nabla_{x}^{2}\right)
V^{3}\left(  x\right)  .
\end{align}
The third-order global heat kernel is
\begin{align}
K_{3}\left(  \tau\right)   &  =\int\left\langle x\left\vert K_{3}\right\vert
x\right\rangle d^{\nu}x\nonumber\\
&  =-\frac{1}{\left(  4\pi\tau\right)  ^{\nu/2}}\tau^{3}\int_{0}^{1}u_{3}%
^{2}du_{3}\int_{0}^{1}u_{2}du_{2}\int_{0}^{1}du_{1}\int d^{\nu}x\nonumber\\
&  \times\exp\left(  \left(  1-u_{1}u_{2}u_{3}\right)  u_{1}u_{2}u_{3}%
\tau\nabla_{x}^{2}+\left(  1-u_{2}u_{3}\right)  u_{2}u_{3}\tau\nabla_{x}%
^{2}\right.  \nonumber\\
&  +\left(  1-u_{3}\right)  u_{3}\tau\nabla_{x}^{2}+2\left(  1-u_{2}%
u_{3}\right)  u_{1}u_{2}u_{3}\tau\nabla_{x}^{2}\nonumber\\
&  +\left.  2\left(  1-u_{3}\right)  u_{1}u_{2}u_{3}\tau\nabla_{x}%
^{2}+2\left(  1-u_{3}\right)  u_{2}u_{3}\tau\nabla_{x}^{2}\right)
V^{3}\left(  x\right)  .
\end{align}

\section{Conclusion \label{Conclusion}}

The heat-kernel method is widely used in physics and mathematics. In
application, there are few heat kernel which can be solved exactly. Thus in
most cases, we need to solve the heat kernel approximately. As a kind of
approximation methods, there are many asymptotic method for solving heat
kernels, such as the Seeley-DeWitt type expansion \cite{vassilevich2003heat}
and the covariant perturbation expansion
\cite{barvinsky1985generalized,barvinsky1987beyond,barvinsky1995one,shore2002local,barvinsky1993covariant,barvinsky1994asymptotic,barvinsky1990covariant2,barvinsky1990covariant3,gusev2009heat,barvinsky2002new,barvinsky2010schwinger}%
.

In literature, the covariant perturbation expansion is only used to calculate
diagonal heat kernels. In this paper, we use the covariant perturbation
expansion to calculate non-diagonal heat kernel.

The non-diagonal heat kernel contains all information of an operator.
Therefore, starting from a non-diagonal heat kernel, we can calculate many
physical quantities, such as effective actions, vacuum energies
\cite{dai2010approach}, spectral counting functions \cite{dai2009number},
thermodynamic quantities, and scattering phase shifts
\cite{li2015heat,pang2012relation}. In the further work, we will calculate
such quantities by the result given by the present paper.


\acknowledgments

We are very indebted to Dr G. Zeitrauman for his encouragement. This work is supported in part by NSF of China under Grant No. 11575125 and No. 11375128.










\providecommand{\href}[2]{#2}\begingroup\raggedright\endgroup 


\begin{thebibliography}{10} 
\bibitem{vassilevich2003heat}
D.~V. Vassilevich, {\it Heat kernel expansion: user's manual},  {\em Physics
  Reports} {\bf 388} (2003), no.~5 279--360.

\bibitem{kirsten2001spectral}
K.~Kirsten, {\em Spectral functions in mathematics and physics}.
\newblock CRC Press, 2001.

\bibitem{davies1990heat}
E.~B. Davies, {\em Heat kernels and spectral theory}, vol.~92.
\newblock Cambridge University Press, 1990.

\bibitem{dai2010approach}
W.-S. Dai and M.~Xie, {\it An approach for the calculation of one-loop
  effective actions, vacuum energies, and spectral counting functions},  {\em
  Journal of High Energy Physics} {\bf 2010} (2010), no.~6 1--29.

\bibitem{mukhanov2007introduction}
V.~Mukhanov and S.~Winitzki, {\em Introduction to quantum effects in gravity}.
\newblock Cambridge University Press, 2007.

\bibitem{dai2003quantum}
W.-S. Dai and M.~Xie, {\it Quantum statistics of ideal gases in confined
  space},  {\em Physics Letters A} {\bf 311} (2003), no.~4 340--346.

\bibitem{dai2004geometry}
W.-S. Dai and M.~Xie, {\it Geometry effects in confined space},  {\em Physical
  Review E} {\bf 70} (2004), no.~1 016103.

\bibitem{dai2009number}
W.-S. Dai and M.~Xie, {\it The number of eigenstates: counting function and
  heat kernel},  {\em Journal of High Energy Physics} {\bf 2009} (2009), no.~02
  033.

\bibitem{li2015heat}
W.-D. Li and W.-S. Dai, {\it Heat-kernel approach for scattering},  {\em The
  European Physical Journal C} {\bf 75} (2015), no.~6.

\bibitem{pang2012relation}
H.~Pang, W.-S. Dai, and M.~Xie, {\it Relation between heat kernel method and
  scattering spectral method},  {\em The European Physical Journal C-Particles
  and Fields} {\bf 72} (2012), no.~5 1--13.

\bibitem{fliegner1994higher}
D.~Fliegner, M.~G. Schmidt, and C.~Schubert, {\it The higher derivative
  expansion of the effective action by the string-inspired method. i},  {\em
  Zeitschrift f{\"u}r Physik C Particles and Fields} {\bf 64} (1994), no.~1
  111--116.

\bibitem{fliegner1998higher}
D.~Fliegner, P.~Haberl, M.~Schmidt, and C.~Schubert, {\it The higher derivative
  expansion of the effective action by the string inspired method, ii},  {\em
  Annals of Physics} {\bf 264} (1998), no.~1 51--74.

\bibitem{nepomechie1985calculating}
R.~I. Nepomechie, {\it Calculating heat kernels},  {\em Physical Review D} {\bf
  31} (1985), no.~12 3291.

\bibitem{culumovic1988calculation}
L.~Culumovic and D.~McKeon, {\it Calculation of off-diagonal elements of the
  heat kernel},  {\em Physical Review D} {\bf 38} (1988), no.~12 3831.

\bibitem{dilkes1996off}
F.~Dilkes and D.~McKeon, {\it Off-diagonal elements of the dewitt expansion
  from the quantum-mechanical path integral},  {\em Physical Review D} {\bf 53}
  (1996), no.~8 4388.

\bibitem{mckeon1991seeley}
D.~McKEON, {\it Seeley-gilkey coefficients for superoperators},  {\em Modern
  Physics Letters A} {\bf 6} (1991), no.~40 3711--3715.

\bibitem{kotani2000albanese}
M.~Kotani and T.~Sunada, {\it Albanese maps and off diagonal long time
  asymptotics for the heat kernel},  {\em Communications in Mathematical
  Physics} {\bf 209} (2000), no.~3 633--670.

\bibitem{groh2011off}
K.~Groh, F.~Saueressig, and O.~Zanusso, {\it Off-diagonal heat-kernel expansion
  and its application to fields with differential constraints},  {\em arXiv
  preprint arXiv:1112.4856} (2011).

\bibitem{barvinsky1985generalized}
A.~Barvinsky and G.~Vilkovisky, {\it The generalized schwinger-dewitt technique
  in gauge theories and quantum gravity},  {\em Physics Reports} {\bf 119}
  (1985), no.~1 1--74.

\bibitem{barvinsky1987beyond}
A.~Barvinsky and G.~Vilkovisky, {\it Beyond the schwinger-dewitt technique:
  Converting loops into trees and in-in currents},  {\em Nuclear Physics B}
  {\bf 282} (1987) 163--188.

\bibitem{barvinsky1995one}
A.~Barvinsky, Y.~V. Gusev, G.~Vilkovisky, and V.~Zhytnikov, {\it The one-loop
  effective action and trace anomaly in four dimensions},  {\em Nuclear Physics
  B} {\bf 439} (1995), no.~3 561--582.

\bibitem{shore2002local}
G.~M. Shore, {\it A local effective action for photon--gravity interactions},
  {\em Nuclear Physics B} {\bf 646} (2002), no.~1 281--300.

\bibitem{barvinsky1993covariant}
A.~Barvinsky, V.~Zhytnikov, Y.~V. Gusev, and G.~Vilkovisky, {\it Covariant
  perturbation theory; 4, third order in the curvature},  tech. rep.,
  P00011539, 1993.

\bibitem{barvinsky1994asymptotic}
A.~Barvinsky, Y.~V. Gusev, G.~Vilkovisky, and V.~Zhytnikov, {\it Asymptotic
  behaviors of the heat kernel in covariant perturbation theory},  {\em Journal
  of Mathematical Physics} {\bf 35} (1994), no.~7 3543--3559.

\bibitem{barvinsky1990covariant2}
A.~Barvinsky and G.~Vilkovisky, {\it Covariant perturbation theory (ii). second
  order in the curvature. general algorithms},  {\em Nuclear Physics B} {\bf
  333} (1990), no.~2 471--511.

\bibitem{barvinsky1990covariant3}
A.~Barvinsky and G.~Vilkovisky, {\it Covariant perturbation theory (iii).
  spectral representations of the third-order form factors},  {\em Nuclear
  Physics B} {\bf 333} (1990), no.~2 512--524.

\bibitem{gusev2009heat}
Y.~V. Gusev, {\it Heat kernel expansion in the covariant perturbation theory},
  {\em Nuclear physics B} {\bf 807} (2009), no.~3 566--590.

\bibitem{barvinsky2002new}
A.~Barvinsky and V.~Mukhanov, {\it New nonlocal effective action},  {\em
  Physical Review D} {\bf 66} (2002), no.~6 065007.

\bibitem{barvinsky2010schwinger}
A.~Barvinsky and D.~Nesterov, {\it Schwinger-dewitt technique for quantum
  effective action in brane induced gravity models},  {\em Physical Review D}
  {\bf 81} (2010), no.~8 085018.

\bibitem{van1998index}
A.~E. Van~de Ven, {\it Index-free heat kernel coefficients},  {\em Classical
  and Quantum Gravity} {\bf 15} (1998), no.~8 2311.
\end{thebibliography}

\end{document}